\documentclass{llncs}
\usepackage{verbatim}
\usepackage{listings}
\usepackage{todonotes}
\usepackage{xcolor}

\definecolor{color:keyword}{rgb}{0.53,0.05,0.05}
\definecolor{color:comment}{rgb}{0.25,0.37,0.75}
\definecolor{color:string}{rgb}{0.87,0.0,0.0}
\usepackage{bold-extra}

\lstdefinelanguage{Jolie}{
morekeywords={
provide,until,OneWay,RequestResponse,new, type, main,define,inputPort,outputPort,init,execution,include,	cset,if,else,csets,interface, throws,global,constants,for,
foreach,while,int,double,raw,void,undefined,string,long,bool,any,single, sequential, concurrent, Jolie, Java, JavaScript, embedded, Location, Protocol, Interfaces, Aggregates, scope, install, cH, comp, throw, this, default, synchronized, nullProcess, false, true},
sensitive=true,
comment=[l]{//},
morecomment=[s]{/*}{*/},
morestring=[b]",
otherkeywords={;,|,:}
}

\lstdefinelanguage{z3}{
morekeywords={
assert, false, true, declare, sort, fun, Bool, String, RegEx, forall, const, define, let, not, implies, iff, and, or},
sensitive=true,
comment=[l]{//},
morecomment=[s]{/*}{*/},
comment=[l]{;},
morestring=[b]",
otherkeywords={|,:}
}

\begin{document}

\title{Refinement types in Jolie}
\titlerunning{}  
%
\author{
Alexander Tchitchigin\inst{1,2}
\and Larisa Safina \inst{1}
\and Manuel Mazzara \inst{1} \\
     Mohamed Elwakil \inst{1} \thanks{Dr. Mohamed Elwakil is on a sabbatical leave from Cairo University, Giza, Egypt.}
\and Fabrizio Montesi \inst{3} \thanks{Supported by CRC (Choreographies for Reliable and efficient Communication software), grant no. DFF-–4005-00304 from the Danish Council for Independent Research.}
\and Victor Rivera \inst{1}
}
\institute{Innopolis university, Innopolis, Russia,\\
\email{\{a.chichigin, l.safina, m.mazzara, m.elwakil, v.rivera\}@innopolis.ru}
\and
Kazan Federal University, Russia \qquad
\email{a.tchichigin@it.kfu.ru}
\and
University of Southern Denmark, Denmark\qquad 
\email{fmontesi@imada.sdu.dk}
}

\authorrunning{} 
%
\tocauthor{}
%

\maketitle              

\begin{abstract}

Jolie is the first language for microservices and it is currently dynamically type checked. This paper considers the opportunity to integrate dynamic and static type checking with the introduction of refinement types, verified via SMT solver. The integration of the two aspects allows a scenario where the static verification of \textit{internal} services and the dynamic verification of (potentially malicious) \textit{external} services cooperates in order to reduce testing effort and enhancing security.

\end{abstract}

\section{Introduction}
``Stringly typed'' is a new antipattern referring to an implementation that needlessly relies on strings, when other options are available. The problem of ``string typing" appears often in service-oriented architecture and microservices on the border between a service and its clients (external interfaces) due to necessity to communicate over text-based protocols (like HTTP) and collaboration with clients written in dynamically-typed languages (like JavaScript). The solution to this problem can be found with refinement types, which are used to statically (or dynamically) check compatibility of a given value and refined type by means of predicates constraining the set of possible values. Though employment of numerical refinements is well-known in programming languages, string refinements are still rare.

In this paper, we introduce a design for extending the Jolie programming language~\cite{MGZ14,jolie:website} and its type system. On top of previous extensions with choice type~\cite{Safina2016} and regular expressions, we introduce here string refinement type and we motivate the reasons for such extension. Section~\ref{jpl} recalls the basic of the Jolie language and its type system while Section~\ref{idea} describes the open problem this paper attacks with clarifying examples. Section~\ref{related-work} discusses related work in the context of using SMT solvers for static typing of refinement types.

\section{Jolie programming language}\label{jpl}

Jolie~\cite{MGZ14} is the first programming language based on the paradigm of microservices~\cite{ms:fowler}: all components are autonomous services that can be deployed independently and operate by running parallel processes, programmed following the workflow approach. Microservices can be composed to obtain, in turn, other microservices.
The language was originally developed in the context of a major formalization effort for workflow and services composition languages, the EU Project SENSORIA~\cite{sensoria}, which spawned many models for reasoning on the composition of services (e.g.,~\cite{LucchiM07,Mazzara11}).
Jolie comes with a formally-specified semantics~\cite{sock,GLMZ09,MC11}; on the more practical side it is inspired by standards for Service-oriented Computing such as WS-BPEL~\cite{bpel}. The combination of theoretical and practical aspects in Jolie enabled its usage in research on correct-by-construction software (see, e.g.,~\cite{PGLMG14,CM13,M13:phd}).




Microservices work together by exchanging messages. In Jolie, messages are structured as trees~\cite{MC11} (a variant of the structures that can be found in XML or JSON).
Communications are type checked at runtime, when messages are sent or received. Type checking of incoming messages is especially relevant, since it mitigates the effect of ill-behaved clients. The work in~\cite{nielsen} presents a first attempt at formalizing a static type checker for the core fragment of Jolie. However, for the time being, the language is still dynamically type checked.

\section{Extension of Jolie Type System}\label{idea}

In~\cite{Safina2016}, the basic type system of Jolie has been extended with type choices.
The work had been then continued with the addition of regular expression types,
a special case of refinement types~\cite{Freeman:1991}.
In refinement types, types are decorated with logical predicates, which further constrain the set of values described by the type and therefore represent the specification of invariant on values. Here, we extend this with the possibility of expressing invariants on string values in form of regular expressions.
%
%
%
%
%
%
%
The integration of static and dynamic analysis allows considering \textit{“internal”} services (native Jolie services) and calls from \textit{“external”} services (potentially developed in other languages) in a complementary way. The first ones can be statically checked while the second ones, which could exhibit malicious behavior, still need a runtime validation.

The key idea behind service-oriented computing, and microservices in particular, is the ability to connect services developed in different programming languages and possibly running on different servers
over standard communication protocols~\cite{ms:fowler}. 
A common use case is the implementation of APIs for Web and mobile applications. In such scenarios, the de-facto standard communication protocol is HTTP(S), combined with standardized data formats (SOAP, JSON, etc.).

HTTP is a text-based protocol, where all data get serialized into strings\footnote{Jolie partially mitigates this aspect with automatic conversion of string serializations to structured data by following the interface definition of the service~\cite{M14}. However, this does not solve the general problem addressed here.}. Moreover, clients of a service (an application or another service) may have been developed in a language that does not support particular datatypes (e.g., JavaScript does not have a datatype for calendar dates or time of day), therefore relying on string representation for internal processing too. The same issue arises with key-value storage systems (e.g., Memcache and Redis), which support only string keys and string values. These factors make string handling an important part of a service application, especially at the boundary with external systems. 

Not all strings are made equal. For example, GUIDs are often used to identify records in a store. GUIDs are represented as strings of hexadecimal digits with a particular structure. 
Currently, developers have to manually check the conformance of received values to the expected format. 

Description of the \textit{shape} of expected string data 
is natural with 
regular expressions. Adding the description of this \textit{shape} to the datatype definition allows the compiler to automatically insert the necessary dynamic checks and statically validate the conformance. 
This is the extension 
of refinement type to string type. 
The same techniques and tools used for static verification of conformance for numerical refinements~\cite{knowles2006sage,dunfield2007unified}
can be used for strings. For the purposes of this paper we will use Z3 SMT solver by Microsoft Research \cite{Z3}, which recently got support for theory of strings and regular expressions in development branch.

\subsection{Example: the news board}

The approach to static checking of string refinements using Z3 SMT solver is illustrated here by a simple example, i.e. a service using refined datatype for GUIDs and the SMT constraints generated for it.

A \textit{news board} is a simple service in charge of retrieving posts composed by a particular user of the system. The service receives user information via HTTP in a string format. String refinement types allow the definition of constraints on user IDs as an alternative to the implementation of the logic checking the constraint inside the posts retrieving operation.

\begin{lstlisting}[language=jolie]
type guid: string("[A-F\\d]{8,8}-[A-F\\d]{4,4}-[A-F\\d]
                      {4,4}-[A-F\\d]{4,4}-[A-F\\d]{12,12}")
\end{lstlisting}

Types for storing user and posts information are also necessary.

\begin{lstlisting}[language=jolie]
type user: void {
  .uid: guid
  .name: string
  .age: int(age>18) }
type post_type: void {
  .pid: guid
  .owner: guid
  .content: string }
type posts: void { .post*: post_type }
\end{lstlisting}

We leave service deployment information out of this paper due to its low relevance to the topic, the full code example can be found in~\cite{smt-constraints:gist}. The behavioral fragment of the \textit{news board} demonstrates the post retrieval for a particular user. To get the information the right user has to be found (\textit{find\_user\_by\_name}) and pass the GUID to \textit{get{\_}all{\_}users{\_}posts}. 

There are two definitions of the operation in the following code fragment: \textit{all{\_}posts{\_}by{\_}user} and \textit{all{\_}posts{\_}by{\_}user2}. In the first one the correct data is passed to \textit{get{\_}all{\_}users{\_}posts}, i.e. \textit{user.uid}; while in the second \textit{user.name} is passed. Without string refinement a problem would here arise. The code is syntactically correct. However, it's semantically incorrect since no information can be retrieved by user's name when user's ID is actually expected.  

\begin{lstlisting}[language=jolie]
main {
  all_posts_by_user (name) {
    find_user_by_name@SelfOut(name)(user);
    get_all_users_posts@SelfOut(user.uid)(posts) };
    
  all_posts_by_user2 (name) { 
      find_user_by_name@SelfOut(name)(user);
      // and here we pass the wrong field!
      get_all_users_posts@SelfOut(user.name)(posts) };
  
  //find_user_by_name definition
  //get_all_users_posts definition }
\end{lstlisting}

Introducing string refinement allows Jolie to have both dynamic and static checking for strings. In case of dynamic checking, the string is verified at runtime when passed to the receiving service. The more interesting case is static checking by means of SMT. Here we present the most essential parts of the encoding, complete example can be found in~\cite{smt-constraints:gist}.

\begin{lstlisting}[language=z3]
; notions of types, terms and typing relation
(declare-sort Type)
(declare-sort Term)
(declare-fun HasType (Term Type) Bool)

; type of strings of a programming language
(declare-fun string () Type)
; translation from Z3 built-in String type
; to our string type and back
(declare-fun BoxString (String) Term)
(declare-fun string-term-val (Term) String)
(assert (forall ((str String))
  (= (string-term-val (BoxString str)) str)))
(assert (forall ((s String))
  (HasType (BoxString s) string)))

; guid type that refines string type
(declare-fun guid () Type)
(define-fun guid-re () (RegEx String)
; the construction of the regular expression is omitted
)
; refinement definition for guid type
(assert (forall ((x Term))
  (iff (HasType x guid)
       (and (HasType x string)
            (str.in.re (string-term-val x) guid-re)))))
; we define type user through it's projections
(declare-fun user () Type)
(declare-fun user.uid (Term) Term)
(declare-fun user.name (Term) Term)
(declare-fun user.age (Term) Term)
; typing rules for projections
(assert (forall ((t Term))
  (implies (HasType t user)
    (and (HasType (user.uid t)  guid)
         (HasType (user.name t) string)
         (HasType (user.age t)  nat)))))

(declare-fun find_user_by_name (Term) Term)
; find_user_by_name : string -> user
(assert (forall ((name Term))
  (implies (HasType name string)
    (HasType (find_user_by_name name) user))))

; type checking for all_posts_by_user
(assert (not (forall ((t Term))
  (implies (HasType t string)
     (HasType (user.uid (find_user_by_name t)) guid)))))
; type checking for all_posts_by_user2
(assert (not (forall ((t Term))
  (implies (HasType t string)
     (HasType (user.name (find_user_by_name t)) guid)))))
\end{lstlisting}

Type checking is based on proving a theorem stating that a function is correctly typed. Technically, the opposite proposition is actually stated and the SMT solver is put in charge of finding a counterexample. A failure in such an attempt leads to the conclusion that the original theorem has be true (proof by contradiction). 

The Z3 solver successfully proves the well-typedness theorem for the correct implementation of \textit{all\_posts\_by\_user}, and fails to disprove the incorrect implementation (\textit{all\_posts\_by\_user2}). Actually, in the second case the proof never terminates. This fact is due to many simplifications to the presented SMT encoding for the sake of clarity and understandability which cause infinite (recursive) generation of Skolem terms. Employment of a more sophisticated encoding for the actual implementation of refinement constraints may mitigate infinite recursion and it is left as future work.

\section{Related work} \label{related-work}

Within the context of functional languages, type-checking of refined types by employing SMT solvers is not new. In~\cite{Bengtson:2011:RTS:1890028.1890031}, the authors present the design and implementation of the F7 enhanced type-checker for the functional language F\# that verifies security properties of cryptographic protocols and access control mechanisms using Z3~\cite{DeMoura:2008:ZES:1792734.1792766}. The SAGE language~\cite{knowles2006sage} employs a hybrid approach~\cite{Flanagan:2006:HTC:1111037.1111059} that performs both static and dynamic type-checking. During compilation time, the Simplify theorem prover~\cite{Detlefs:2005:STP:1066100.1066102} is used to check refinement types. If Simplify is not able to decide a particular subtyping relation, a proper type cast is inserted in the code and it is checked at runtime. If the type cast fails during runtime, this particular subtyping relation is inserted in a database of known failed casts. In contrast to checking syntactic subtyping as in F7 and SAGE, the authors of~\cite{Bierman:2010:SSS:1863543.1863560}, introduce semantic subtyping checking for a subset of the M language~\cite{PowerQuery} using the Z3 SMT solver. 


\bibliographystyle{plain}
\bibliography{biblio}

\begin{thebibliography}{10}

\bibitem{sensoria}
{EU Project SENSORIA. Accessed February 2016}.
\newblock \url{http://www.sensoria-ist.eu/}.

\bibitem{smt-constraints:gist}
{Gist of SMT constraints for the example. Accessed February 2016}.
\newblock \url{https://gist.github.com/gabriel-fallen/a04c33860e2157201fa8}.

\bibitem{jolie:website}
{Jolie Programming Language. Accessed February 2016.}
\newblock \url{http://www.jolie-lang.org/}.

\bibitem{bpel}
{WS-BPEL} {OASIS} {W}eb {S}ervices {B}usiness {P}rocess {E}xecution
  {L}anguage.accessed february 2016.
\newblock
  \url{http://docs.oasis-open.org/wsbpel/2.0/wsbpel-specification-draft.html}.

\bibitem{PowerQuery}
Power query (informally known as \"m\") formula reference.
\newblock Technical Report, aug 2015.

\bibitem{Z3}
Microsoft Research.Accessed~February 2016.
\newblock {Z3}.
\newblock \url{https://github.com/Z3Prover/z3}.

\bibitem{Bengtson:2011:RTS:1890028.1890031}
Jesper Bengtson, Karthikeyan Bhargavan, C{\'e}dric Fournet, Andrew~D. Gordon,
  and Sergio Maffeis.
\newblock Refinement types for secure implementations.
\newblock {\em ACM Trans. Program. Lang. Syst.}, 33(2):8:1--8:45, February
  2011.

\bibitem{Bierman:2010:SSS:1863543.1863560}
Gavin~M. Bierman, Andrew~D. Gordon, C\u{a}t\u{a}lin Hri\c{t}cu, and David
  Langworthy.
\newblock Semantic subtyping with an smt solver.
\newblock In {\em Proceedings of the 15th ACM SIGPLAN International Conference
  on Functional Programming}, ICFP '10, pages 105--116, New York, NY, USA,
  2010. ACM.

\bibitem{CM13}
Marco Carbone and Fabrizio Montesi.
\newblock Deadlock-freedom-by-design: multiparty asynchronous global
  programming.
\newblock In {\em POPL}, pages 263--274, 2013.

\bibitem{DeMoura:2008:ZES:1792734.1792766}
Leonardo De~Moura and Nikolaj Bj{\o}rner.
\newblock Z3: An efficient smt solver.
\newblock In {\em Proc. of 14th International Conference on Tools and
  Algorithms for the Construction and Analysis of Systems}, TACAS'08/ETAPS'08,
  pages 337--340, Berlin, Heidelberg, 2008. Springer-Verlag.

\bibitem{Detlefs:2005:STP:1066100.1066102}
David Detlefs, Greg Nelson, and James~B. Saxe.
\newblock Simplify: A theorem prover for program checking.
\newblock {\em J. ACM}, 52(3):365--473, May 2005.

\bibitem{dunfield2007unified}
Joshua Dunfield.
\newblock {\em A unified system of type refinements}.
\newblock PhD thesis, Air Force Research Laboratory, 2007.

\bibitem{Flanagan:2006:HTC:1111037.1111059}
Cormac Flanagan.
\newblock Hybrid type checking.
\newblock In {\em Conference Record of the 33rd ACM SIGPLAN-SIGACT Symposium on
  Principles of Programming Languages}, POPL '06, pages 245--256, New York, NY,
  USA, 2006. ACM.

\bibitem{Freeman:1991}
Tim Freeman and Frank Pfenning.
\newblock Refinement types for ml.
\newblock In {\em Proceedings of the ACM SIGPLAN 1991 Conference on Programming
  Language Design and Implementation}, PLDI '91, pages 268--277. ACM, 1991.

\bibitem{GLMZ09}
Claudio Guidi, Ivan Lanese, Fabrizio Montesi, and Gianluigi Zavattaro.
\newblock Dynamic error handling in service oriented applications.
\newblock {\em Fundam. Inform.}, 95(1):73--102, 2009.

\bibitem{sock}
Claudio Guidi, Roberto Lucchi, Gianluigi Zavattaro, Nadia Busi, and Roberto
  Gorrieri.
\newblock Sock: a calculus for service oriented computing.
\newblock In {\em In ICSOC, volume 4294 of LNCS}, pages 327--338. Springer,
  2006.

\bibitem{ms:fowler}
Martin~Fowler James~Lewis.
\newblock {Microservices: a definition of this new architectural term. Accessed
  February 2016.}
\newblock \url{http://martinfowler.com/articles/microservices.htm}.

\bibitem{knowles2006sage}
Kenneth Knowles, Aaron Tomb, Jessica Gronski, Stephen~N Freund, and Cormac
  Flanagan.
\newblock Sage: Unified hybrid checking for first-class types, general
  refinement types, and dynamic (extended report), 2006.

\bibitem{LucchiM07}
Roberto Lucchi and Manuel Mazzara.
\newblock A pi-calculus based semantics for {WS-BPEL}.
\newblock {\em J. Log. Algebr. Program.}, 70(1):96--118, 2007.

\bibitem{Mazzara11}
Manuel Mazzara, Faisal Abouzaid, Nicola Dragoni, and Anirban Bhattacharyya.
\newblock Toward design, modelling and analysis of dynamic workflow
  reconfigurations - {A} process algebra perspective.
\newblock In {\em Web Services and Formal Methods - 8th International Workshop,
  {WS-FM}}, pages 64--78, 2011.

\bibitem{M13:phd}
Fabrizio Montesi.
\newblock {\em Choreographic Programming}.
\newblock Ph.{D}. thesis, IT University of Copenhagen, 2013.
\newblock \url{http://fabriziomontesi.com/files/m13\_phdthesis.pdf}.

\bibitem{M14}
Fabrizio Montesi.
\newblock Process-aware web programming with jolie.
\newblock volume abs/1410.3712, 2014.

\bibitem{MC11}
Fabrizio Montesi and Marco Carbone.
\newblock {Programming Services with Correlation Sets}.
\newblock In {\em Proc. of Service-Oriented Computing - 9th International
  Conference, {ICSOC}}, pages 125--141, 2011.

\bibitem{MGZ14}
Fabrizio Montesi, Claudio Guidi, and Gianluigi Zavattaro.
\newblock Service-oriented programming with jolie.
\newblock In {\em Web Services Foundations}, pages 81--107. 2014.

\bibitem{nielsen}
J.~M. Nielsen.
\newblock {A Type System for the Jolie Language}.
\newblock Master's thesis, Technical University of Denmark, 2013.

\bibitem{PGLMG14}
Mila~Dalla Preda, Saverio Giallorenzo, Ivan Lanese, Jacopo Mauro, and Maurizio
  Gabbrielli.
\newblock {AIOCJ:} {A} choreographic framework for safe adaptive distributed
  applications.
\newblock In {\em Software Language Engineering - 7th International Conference,
  {SLE} 2014, V{\"{a}}ster{\aa}s, Sweden, September 15-16, 2014. Proceedings},
  pages 161--170, 2014.

\bibitem{Safina2016}
Larisa Safina, Manuel Mazzara, Fabrizio Montesi, and Victor Rivera.
\newblock Data-driven workflows for microservices (genericity in jolie).
\newblock In {\em Proc. of The 30th IEEE International Conference on Advanced
  Information Networking and Applications (AINA), 2016}.

\end{thebibliography}

\end{document}